\documentclass[12pt]{article}
\pdfoutput=1
\usepackage{graphicx}
\usepackage{hyperref}

\newcommand\pubnumber{SNSN-323-63}
\newcommand\pubdate{\today}

\def\institute{Karlsruher Institut f\"ur Technologie (KIT)\\
  Institut f\"ur Experimentelle Kernphysik\\
  Hermann-von-Helmholtz-Platz 1\\
  76344 Eggenstein-Leopoldshafen, Germany
}

\def\Title#1{\begin{center} {\Large #1 } \end{center}}
\def\Author#1{\begin{center}{ \sc #1} \end{center}}
\def\Address#1{\begin{center}{ \it #1} \end{center}}

\newcommand\pubblock{\rightline{\begin{tabular}{l} \pubnumber\\
         \pubdate  \end{tabular}}}
\newenvironment{Abstract}{\begin{quotation}  }{\end{quotation}}
\newenvironment{Presented}{\begin{quotation} \begin{center} 
             PRESENTED AT\end{center}\bigskip 
      \begin{center}\begin{large}}{\end{large}\end{center} \end{quotation}}

\usepackage{amsmath}
\usepackage{xspace}

\def\beq{\begin{equation}}
\def\eeq#1{\label{#1}\end{equation}}
\def\eeqn{\end{equation}}

\def\beqa{\begin{eqnarray}}
\def\eeqa#1{\label{#1}\end{eqnarray}}
\def\eeqan{\end{eqnarray}}

\let\bar=\overbar

\def\Dslash{\not{\hbox{\kern-4pt $D$}}}
\def\dslash{\not{\hbox{\kern-2pt $\del$}}}

\def\msb{{\bar{\ssstyle M \kern -1pt S}}}

\newcommand{\yt}{\ensuremath{y_{\text{t}}}\xspace}
\newcommand{\kt}{\ensuremath{\kappa_{\text{t}}}\xspace}
\newcommand{\kV}{\ensuremath{\kappa_{\text{V}}}\xspace}
\newcommand{\ttH}{\ensuremath{\text{t}\bar{\text{t}}\text{H}}\xspace}
\newcommand{\tH}{\ensuremath{\text{t}\text{H}}\xspace}
\newcommand{\Htobb}{\ensuremath{\text{H}\rightarrow\text{b}\bar{\text{b}}}\xspace}
\newcommand{\ttHbb}{\ensuremath{\text{t}\bar{\text{t}}\text{H}(\text{b}\bar{\text{b}})}\xspace}
\newcommand{\tHbb}{\ensuremath{\text{t}\text{H}(\text{b}\bar{\text{b}})}\xspace}

\newcommand{\ttbar}{\ensuremath{\text{t}\bar{\text{t}}}\xspace}
\newcommand{\ttjets}{\ensuremath{\text{t}\bar{\text{t}}+\text{jets}}\xspace}
\newcommand{\ttlf}{\ensuremath{\text{t}\bar{\text{t}}+\text{LF}}\xspace}
\newcommand{\tthf}{\ensuremath{\text{t}\bar{\text{t}}+\text{HF}}\xspace}
\newcommand{\ttbb}{\ensuremath{\text{t}\bar{\text{t}}+\text{b}\bar{\text{b}}}\xspace}

\newcommand{\fbinv}{\ensuremath{\,\text{fb}^{-1}}\xspace}
\newcommand{\pb}{\ensuremath{\,\text{pb}}\xspace}
\newcommand{\fb}{\ensuremath{\,\text{fb}}\xspace}

\newcommand{\pt}{\ensuremath{p_{\text{T}}}\xspace}
\newcommand{\gev}{\ensuremath{\,\text{Ge}\kern-0.06667em\text{V}}\xspace}
\newcommand{\tev}{\ensuremath{\,\text{Te}\kern-0.06667em\text{V}}\xspace}

\begin{document}
\begin{titlepage}
\pubblock

\vfill
\Title{Searches for \ttH and \tH with \Htobb}
\vfill
\Author{Matthias Schr\"oder\\ on behalf of the ATLAS and CMS Collaborations}
\Address{\institute}
\vfill
\begin{Abstract}
\noindent The associated production of a Higgs boson with a top quark-antiquark pair (\ttH production) or with a single top quark (\tH production) allows a direct measurement of the top-Higgs-Yukawa coupling with minimal model dependence.
In this article, recent results of searches for \ttH and \tH production in the \Htobb channel performed by the ATLAS and CMS experiments are reviewed.
The analyses use pp collision data collected at a centre-of-mass energy of 13\tev corresponding to an integrated luminosity of up to 13.2\fbinv.
\end{Abstract}
\vfill
\begin{Presented}
$9^{th}$ International Workshop on Top Quark Physics\\
Olomouc, Czech Republic,  September 19--23, 2016
\end{Presented}
\vfill
\end{titlepage}
\def\thefootnote{\fnsymbol{footnote}}
\setcounter{footnote}{0}

\section{Introduction}

The coupling of the Higgs (H) boson to the top (t) quark is of particular interest.
In the Standard Model (SM), it is of Yukawa type with a strength \yt proportional to the t-quark mass, and hence, it is exceptionally large.
Thus, \yt contributes dominantly to various loop processes both in the SM and also in models of new physics.
The current value of \yt is dominated by indirect constraints derived from measurements of the gluon-gluon fusion H-boson production and the \mbox{$\text{H}\rightarrow\gamma\gamma$} decay rate and depends on the assumption of absence of new particle contributions to the loop amplitudes.

\begin{figure}[htb]
  \centering
  \begin{tabular}{ccc}
    \includegraphics[height=0.14\textheight]{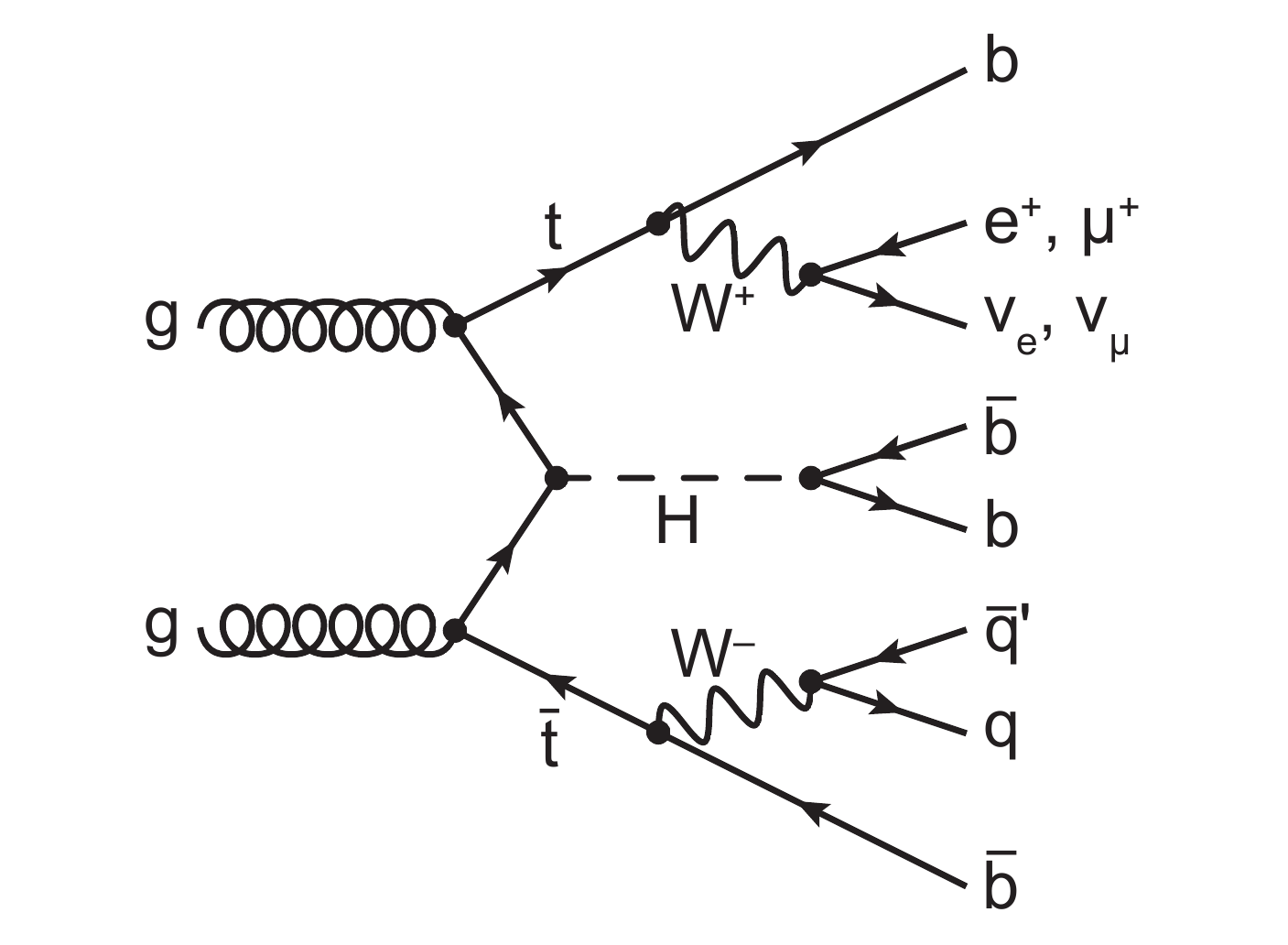} &
    \includegraphics[height=0.14\textheight]{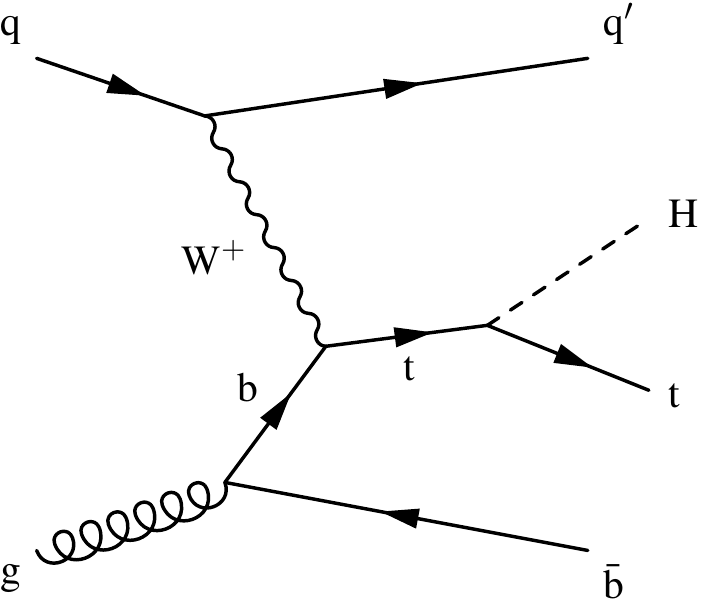} &
    \includegraphics[height=0.14\textheight]{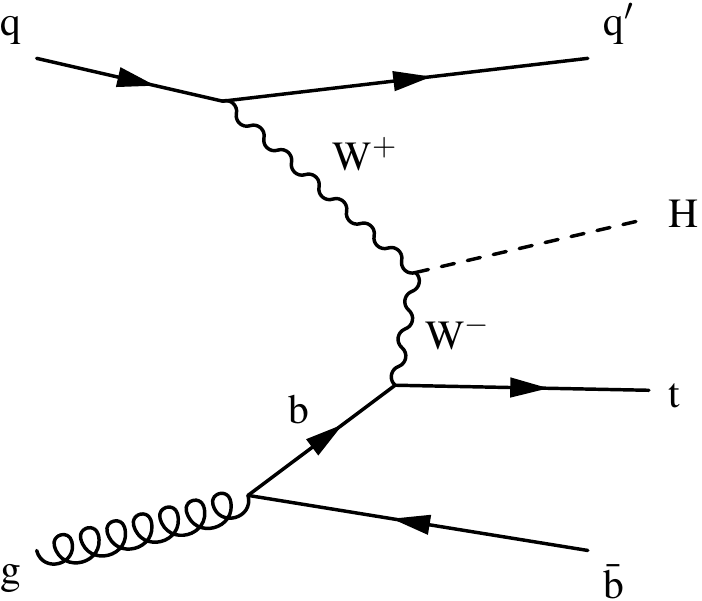} \\
  \end{tabular}
  \caption{
    Example leading-order Feynman diagrams contributing to the lepton+jets- channel \ttHbb (left) and the t-channel \tHbb production (centre, right)~\cite{bib:ttH:cms,bib:tH}.
  }
  \label{fig:diagrams}
\end{figure}

The associated production of a H boson with a t quark-antiquark pair (\ttH production) or with a single t quark (\tH production), on the other hand, allows a direct measurement of \yt with minimal model dependence, cf.\ Fig.~\ref{fig:diagrams}.
However, the SM cross section of both processes is relatively small with approximately 0.5\pb and 90\fb at 13\tev centre-of-mass energy, respectively, making this a difficult measurement.
The bottom (b) quark-antiquark final state of the H boson benefits from a large branching ratio.
At the same time, the relatively poor jet-energy resolution and the huge combinatorial uncertainty in the event reconstruction require to use multivariate analysis methods to discriminate signal from background processes, where the signal cross section is determined with a fit of the discriminant distributions to the data.

\section{Searches for \ttHbb production}
Both CMS~\cite{bib:cms} and ATLAS~\cite{bib:atlas} have performed searches for \ttH, \Htobb, production in the dilepton and lepton+jets final states of the \ttbar system at 13\tev centre-of-mass energy: CMS has published the first analysis at 13\tev using 2.7\fbinv of data collected in 2015~\cite{bib:ttH:cms}, ATLAS has published an analysis using 13.2\fbinv of 2015 and 2016 data~\cite{bib:ttH:atlas}.

Events are generally selected by requiring in the dilepton channel 2 isolated, oppositely-charged leptons (electrons or muons) and $\geq3$ jets, $\geq2$ of which are identified as coming from b quarks (b-tagged), and in the lepton+jets channel 1 isolated lepton and $\geq4$ jets, $\geq2$ of which are b-tagged.
Additional, channel-dependent criteria are applied, such as a Z-boson-mass veto in same-flavour dilepton events.

\begin{figure}[t]
  \centering
  \begin{tabular}{cc}
    \includegraphics[width=0.48\textwidth]{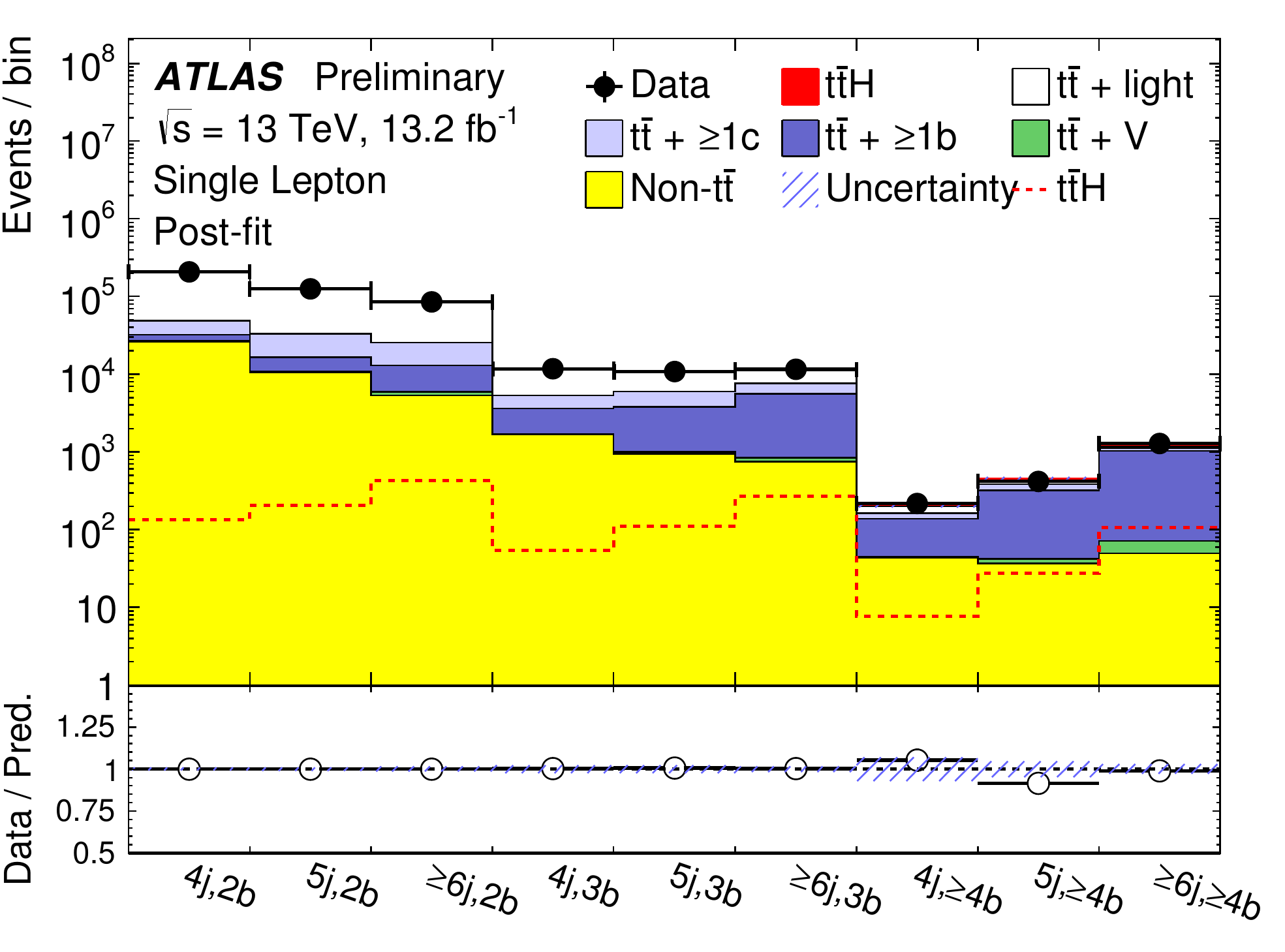} &
    \includegraphics[width=0.38\textwidth]{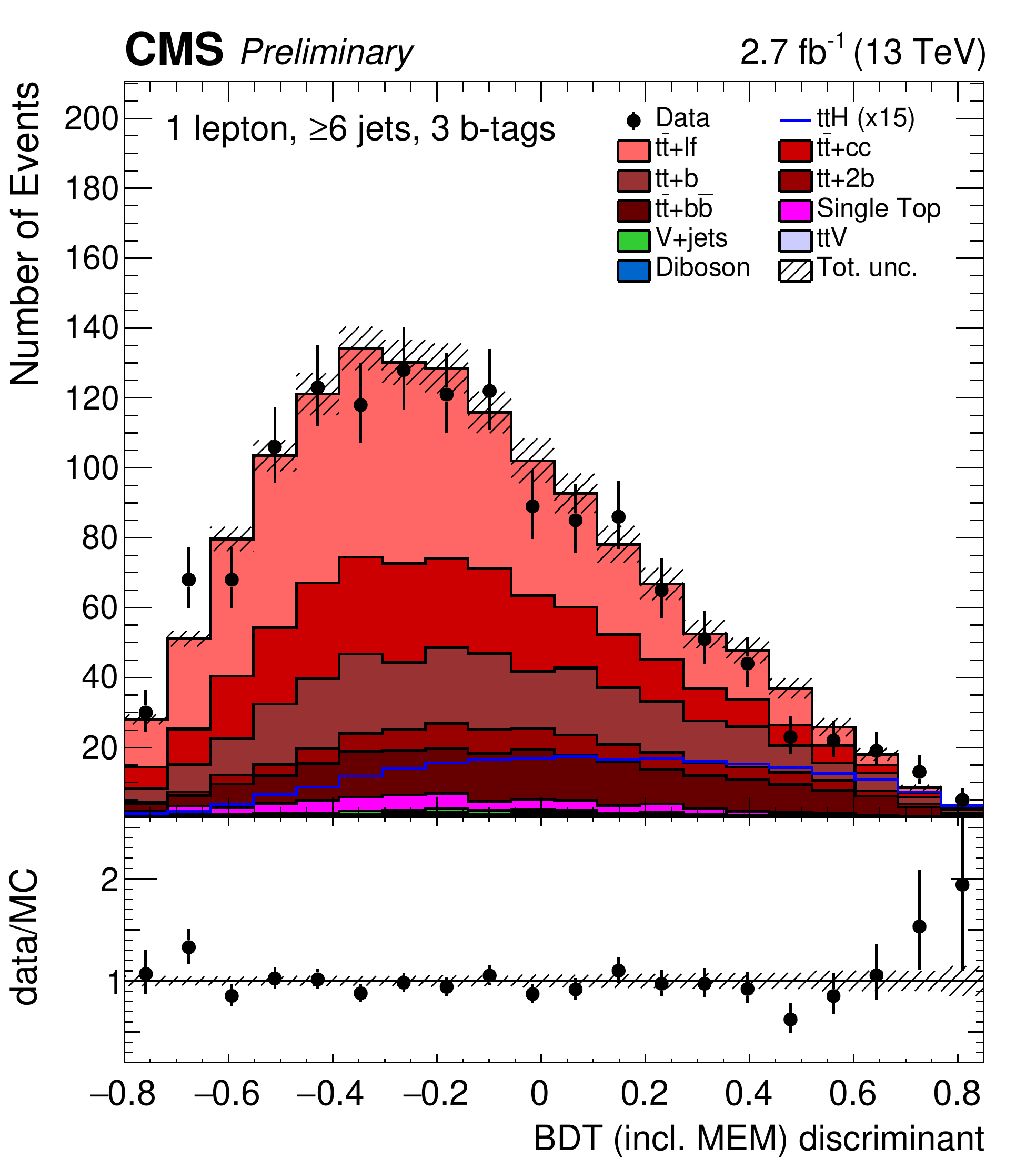} \\
\end{tabular}
  \caption{Predicted and observed event yields per category in the ATLAS analysis~\cite{bib:ttH:atlas} (left) and BDT output in one category of the CMS analysis~\cite{bib:ttH:cms} (right) in the lepton+jets channel, after the fit to the data.
    The \ttjets background is divided by the flavour of the additional jets, and the \ttH contribution, normalised to the best-fit value (left) and 15 times the SM expectation (right), is superimposed.
  }
  \label{fig:ttH:categories}
\end{figure}

Subsequently, the selected events are further divided into mutually exclusive categories in jet and b-tag multiplicity.
Signal events are expected to contribute particularly in the categories with higher multiplicities, cf.\ Fig.~\ref{fig:diagrams} (left).
The SM background consists almost entirely of \ttjets events.
The additional jets stem either from gluons and light-flavour quarks (\ttlf) or from heavy-flavour quarks (\tthf), such as \ttbb events, which represent an irreducible background.
The background composition varies between the categories, cf.\ Fig.~\ref{fig:ttH:categories} (left) as an example.
This is exploited in the final fit, which is performed simultaneously across all categories to constrain the uncertainties of the different processes.
Still, especially the \tthf processes are difficult to model, and the associated uncertainties limit the overall sensitivity.

For the CMS analysis, the \ttjets background is modelled using {\sc Powheg}, interfaced with {\sc Pythia8} with the CUETP8M1 tune for parton showering.
The cross section is normalised to the next-to-next-leading order (NNLO) calculation with resummation to next-to-next-to-leading-logarithmic (NNLL) accuracy using the NNPDF3.0 PDF set.
Events are further separated based on the flavour, defined by the hadron content, of the additional jets within acceptance that do not originate from the t-quark decays, considering also the case that two close-by B hadrons form a single jet.
A rate uncertainty of 50\% is assigned to each of these \tthf processes.

\begin{figure}[htbp]
  \centering
  \begin{tabular}{cc}
    \includegraphics[width=0.4\textwidth]{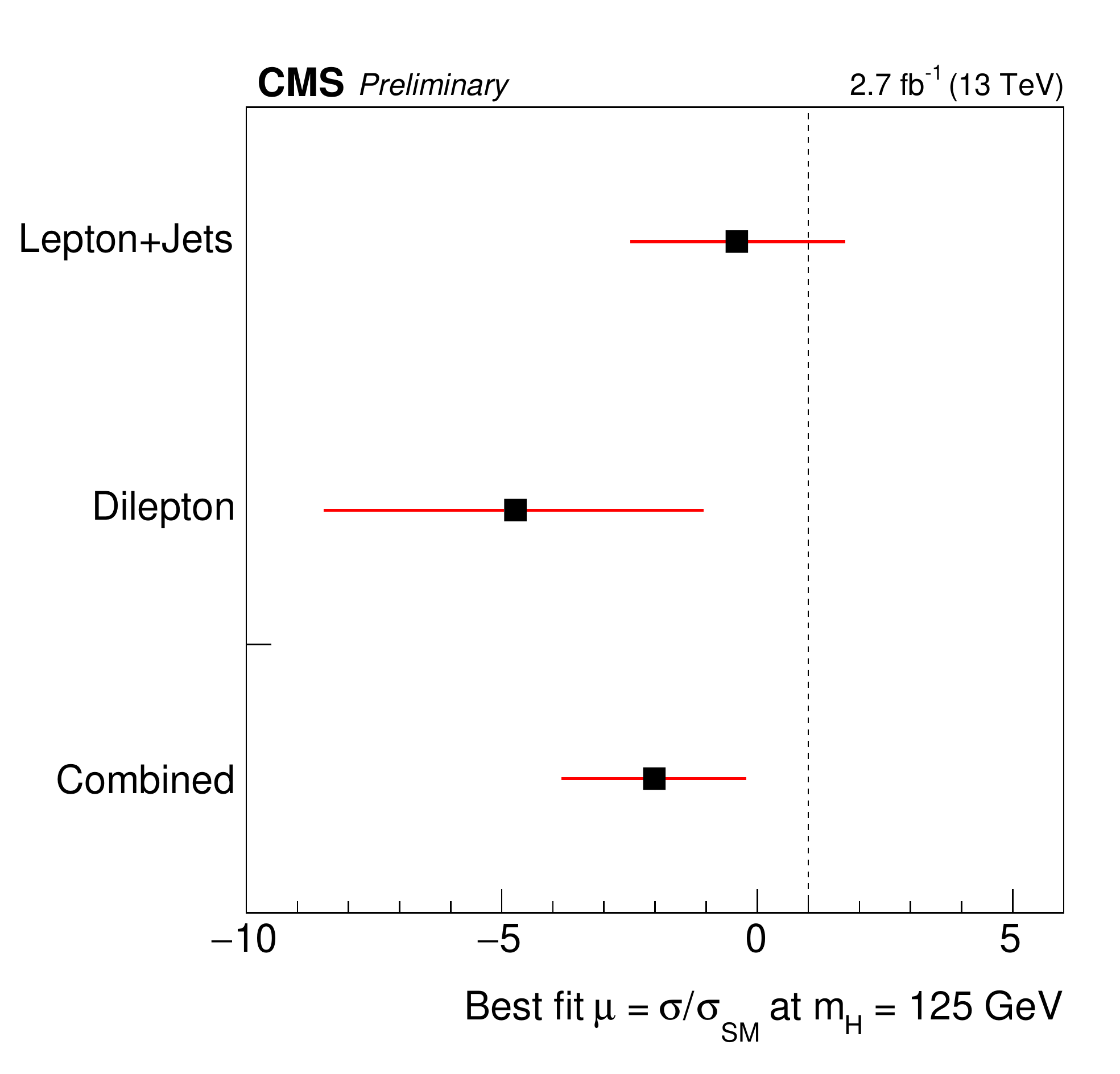} &
    \includegraphics[width=0.5\textwidth]{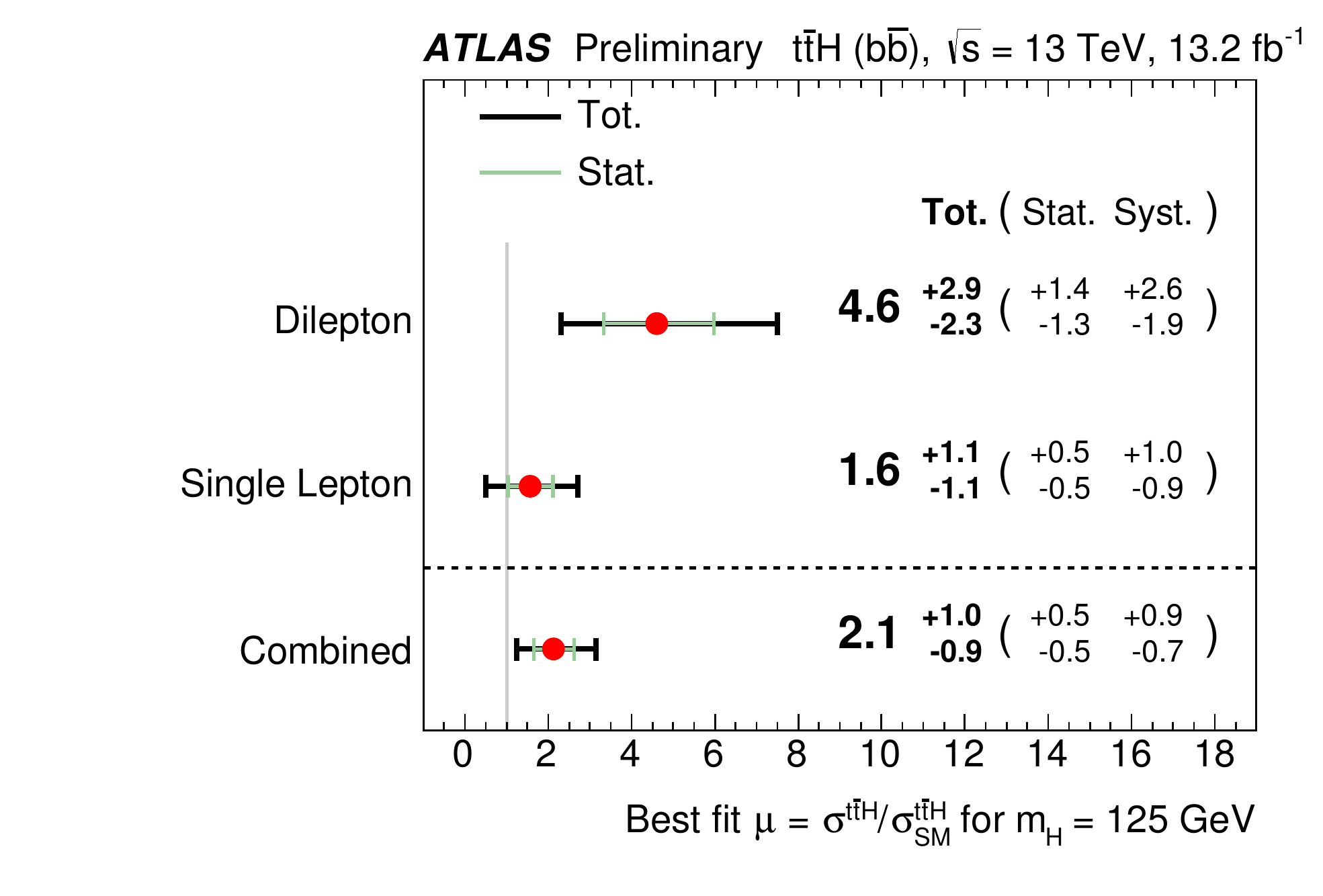} \\
\end{tabular}
  \caption{Best-fit values of the signal-strength modifier $\mu$ with their one standard deviation confidence intervals
    of the CMS~\cite{bib:ttH:cms} (left) and ATLAS~\cite{bib:ttH:atlas} (right) analyses.}
  \label{fig:ttH:results}
\end{figure}

For the ATLAS analysis, the \ttjets background is modelled using {\sc Powheg} and {\sc Pythia6} with the Perugia2012 tune, reweighted to match the t-quark and \ttbar-system \pt spectra of the NNLO calculation, including resummation at NNLL accuracy.
Events with $\geq1$ additional b jets are reweighted to a {\sc Sherpa}+{\sc OpenLoops} prediction at next-to-leading order, using the CT10 4-flavour-scheme PDF set.
Uncertainties on the \tthf processes are assigned based on the differences observed to a {\sc Madgraph5} aMC@NLO prediction, and the overall normalisation of the contributions from \ttbar+$\,\geq1$ b or c jets events is left freely floating in the final fit.

Both analyses employ multivariate methods to combine in each category the information of several variables, such as kinematic properties or invariant masses of combinations of jets and leptons, into a final discriminating variable, cf.\ Fig.~\ref{fig:ttH:categories} (right).
The signal cross section $\sigma$ is determined with a binned fit of the background and signal discriminant distributions to the data, where the uncertainties, which affect the rate and the shape of the distributions, are taken into account via nuisance parameters.

At CMS, boosted decision trees (BDTs) and, in several categories, also a matrix-element-method (MEM) classifier are used.
The latter is a likelihood of the event kinematics under the signal or background hypothesis, taking into account response and acceptance effects, which is constructed to separate against the important \ttbb background.
Depending on the category, the MEM variable is an input to the BDT or events are further separated into two sub-categories with low and high BDT-output and the MEM is used as final discriminant in each sub-category.
In a novel approach, also dedicated techniques are applied to reconstruct events in which the H boson and the hadronically decaying t quark are fairly boosted, resulting in reduced combinatorics in the jet assignment and thus a better event reconstruction efficiency.

At ATLAS, a two-staged multivariate approach is used in the signal-enriched categories.
A BDT is trained to assign the jets to the partons from the H-boson and t-quark decays under the signal hypothesis.
Based on this event reconstruction, additional separating variables, such as the invariant mass of the b jets from the H-boson decay, are computed.
They are used together with reconstruction-independent variables as input to a classification BDT or an artificial neural network that separate signal from background events. 
In the signal-depleted categories in the lepton+jets channel (dilepton channel) the scalar sum of the jet (and lepton) \pt is used, which aims at constraining the systematic uncertainties of the background model.

CMS obtains a signal strength \mbox{$\mu = \sigma/\sigma_{\text{SM}}$} relative to the SM expectation of \mbox{$\mu = -2.0^{+1.8}_{-1.8}$} with 2.7\fbinv of data
and ATLAS of \mbox{$\mu = 2.1^{+1.0}_{-0.9}$} with 13.2\fbinv of data, cf.\ Fig.~\ref{fig:ttH:results}, which are compatible with the SM expectation within 1.7 standard deviations.
These correspond to observed (expected) upper limits on $\mu$ at the 95\% confidence level (C.L.) of 2.6 \mbox{($3.6^{+1.6}_{-1.1}$)} for CMS and 4.0 \mbox{($1.9^{+0.9}_{-0.5}$)} for ATLAS.

\section{Search for \tHbb production}
At leading order, \tH production occurs predominantly via t-channel and associated tW production.
In both cases, the H boson can be emitted either from the t quark or the intermediate W boson, cf.\ Fig.~\ref{fig:diagrams}.
The amplitudes of both contributions interfere depending on the coupling of the H boson to the t quark and to the W boson, expressed hereafter as coupling strengths \kt and \kV relative to the SM expectation, respectively.
Hence, \tH production is sensitive to both the magnitude and the sign of \yt.
The interference is destructive in the SM,
but can in general also be constructive resulting in an enhanced \tH production cross section, e.g.\ by a factor ten for \mbox{$\kt=-1$} and \mbox{$\kV=+1$} (inverted top coupling scenario, ITC).

CMS has performed a search for \tH production in the \Htobb final state with a leptonically decaying
t quark, using 2.3\fbinv of data at 13\tev.
Events are selected requiring 1 isolated lepton and $\geq3$ b-tagged jets, targeting the b quarks from the H boson and t quark decay.
Events are further divided into two exclusive signal regions, one with 3 and one with 4 b-tagged jets as well as 1 additional non b-tagged jet in each case, targeting the fourth b jet, which often falls out of acceptance, and the additional light-flavour jet, cf.\ Fig.~\ref{fig:diagrams}.
The dominant SM background in both signal regions stems from \ttjets production, which is modelled as for the CMS \ttH search.

\begin{figure}[htb]
  \centering
  \begin{tabular}{cc}
    \includegraphics[width=0.48\textwidth]{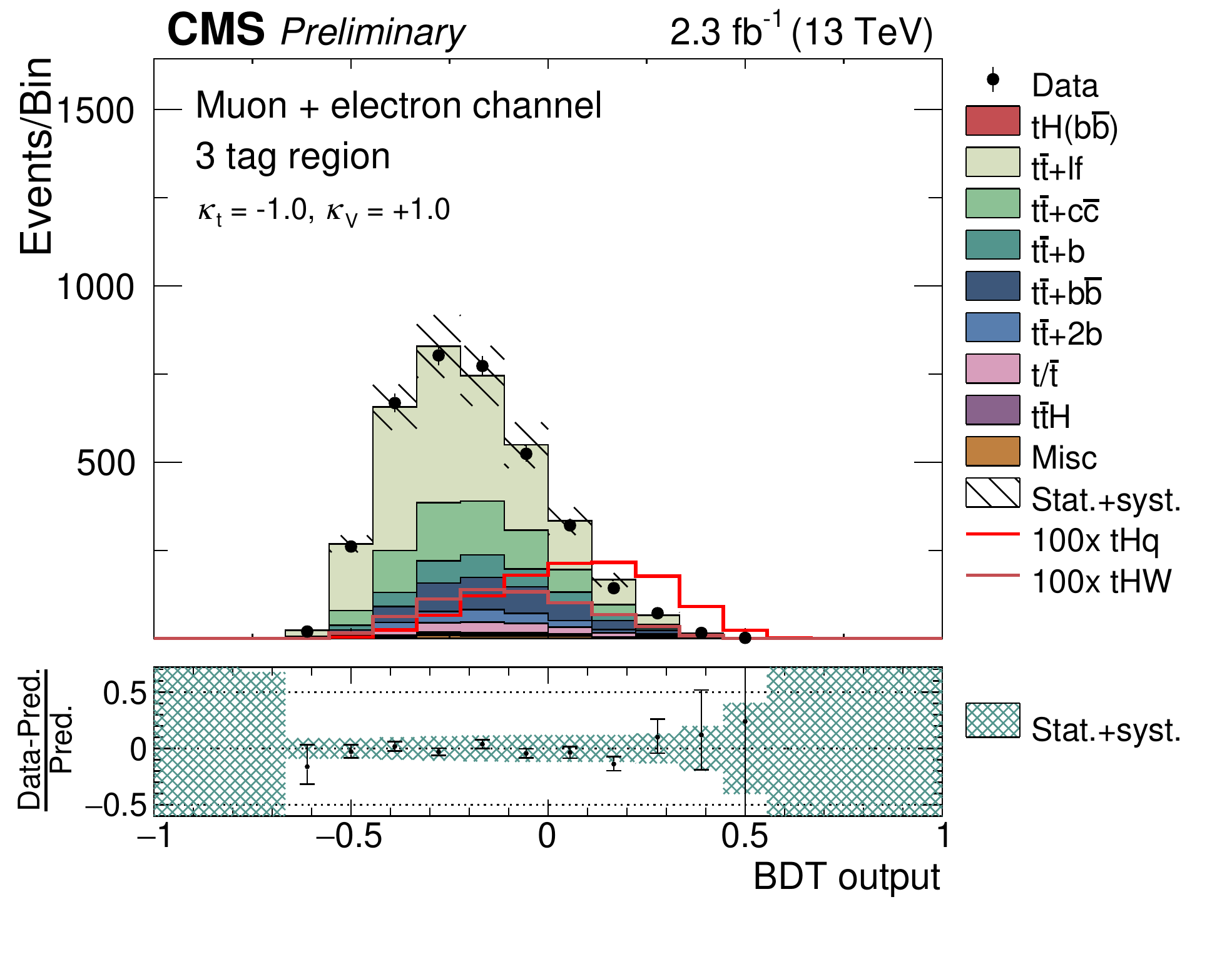} &
    \includegraphics[width=0.42\textwidth]{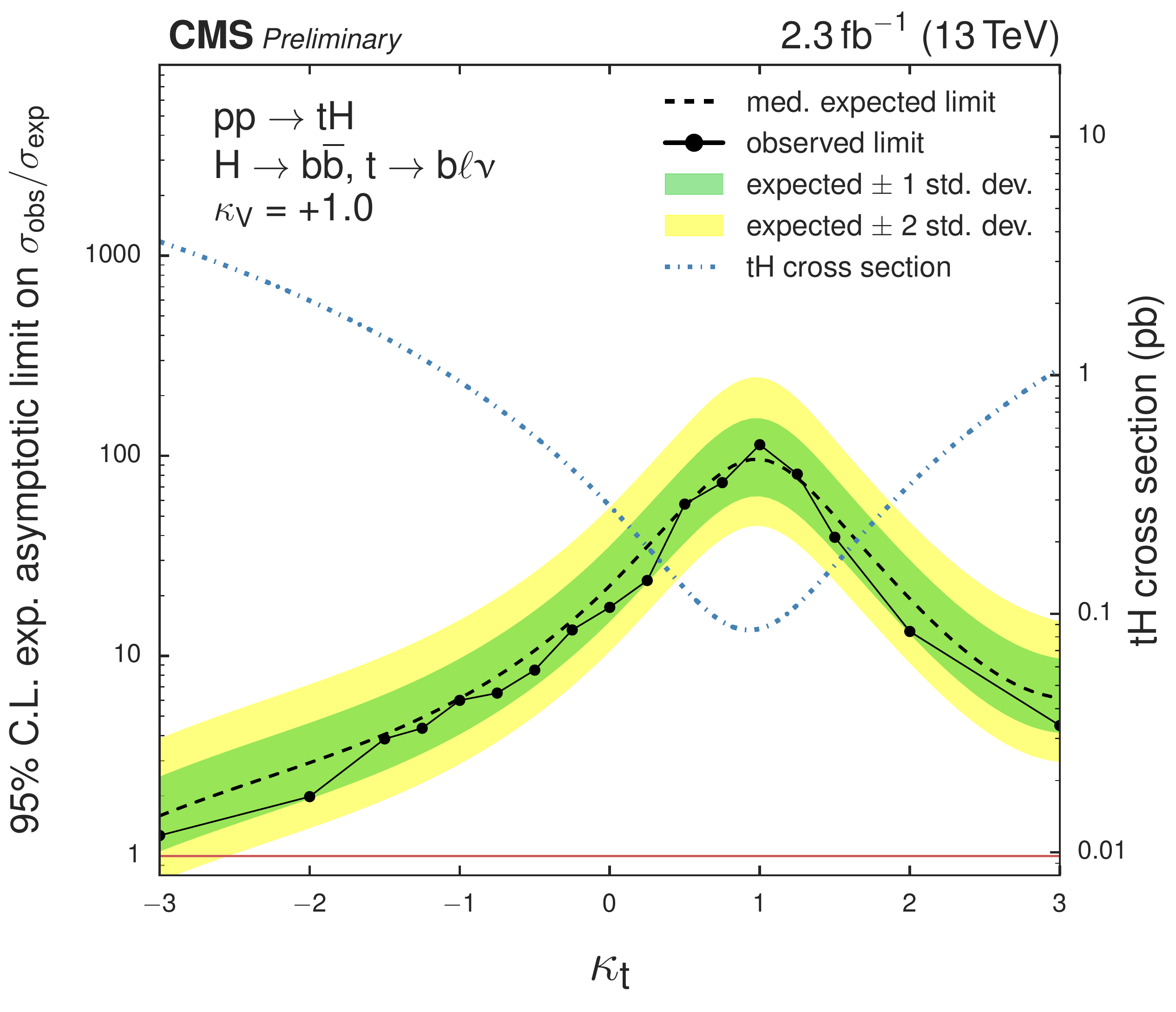} \\
  \end{tabular}
  \caption{Classification BDT output for the ITC case (left), after the fit to the data.
    The \ttjets background is divided by the flavour of the additional jets, and the \tH contributions in both considered production channels, normalised to 100 times the expectation, are superimposed.
    Also shown are the observed and expected upper limits on the signal production cross section as a function of \kt for \mbox{$\kV=+1$} (right)~\cite{bib:tH}.
  }
  \label{fig:tH}
\end{figure}

Events are reconstructed under both the signal and the \ttjets hypothesis, assigning the jets to the final-state quarks depending on the output of two dedicated BDTs that take into account b-tagging and jet kinematic information.
Based on the specific event reconstruction, separating variables, such as the \pt of the H boson, are computed.
These, together with reconstruction-independent variables, are used as input to a classification BDT that separates signal and background, cf.\ Fig.~\ref{fig:tH} (left).

Limits on the signal cross section are obtained for 51 different points in the $(\kt,\kV)$ parameter space by fitting the corresponding BDT output distributions to the data, simultaneously in the two signal regions.
Since the kinematic properties of the signal events depend on \kt and \kV, dedicated BDT trainings are performed for each point.
The results for \mbox{$\kV=+1$} are shown in Fig.~\ref{fig:tH} (right), further results for  \mbox{$\kV=+0.5$} and +1.5 have been derived~\cite{bib:tH}.
For the SM and the ITC scenario, observed (expected) upper limits at 95\% C.L.\ of 113.7 (\mbox{$98.6^{+60.6}_{-34.6}$}) times the SM expectation and 6.0 (\mbox{$6.4^{+3.7}_{-2.2}$}) times the ITC expectation, respectively, are obtained.

\section{Summary}
First searches by CMS and ATLAS for \ttHbb production at 13\tev result in a signal strength of \mbox{$\mu = -2.0^{+1.8}_{-1.8}$} with 2.7\fbinv and \mbox{$\mu = 2.1^{+1.0}_{-0.9}$} with 13.2\fbinv of data.
A 95\% C.L. upper limit on \tHbb production with inverted couplings \mbox{$\kt=-1$} of 6.0 times the expectation is observed by CMS using 2.3\fbinv of data.
The achieved sensitivities are close to or surpass the ones at 8\tev with only a fraction of the data.

\begin{footnotesize}
  
\end{footnotesize}
 
\end{document}